Exoplanets beyond the Conservative Habitable Zone: II. Occurrence

Amri Wandel[1]

[1] Racah Inst. of Physics, The Hebrew University of Jerusalem

Corresponding author, e-mail: amri@huji.ac.il

**Abstract**

We demonstrate that the extension of the Habitable Zone (HZ) due to the presence of liquid water on the night side of tidally locked planets, modelled in this and earlier works, significantly increases the number of potentially habitable planets. We calculate the occurrence of habitable planets orbiting M-, K-, and G-dwarf stars within the conservative and extended HZ, beyond the inner and outer boundaries of the conservative HZ. Integrating over the phase space in the HZ diagram and normalizing our calculation to relatively recent analyses of the Kepler data, we show that potentially habitable planets may be as much as 50 times more abundant than in the lower estimate, limited to the conservative HZ of G-type stars only. For an intermediate heat transport rate on tidally locked planets, we find that the extended HZ could imply more than one habitable planet per star, and hundreds of habitable planets within 10 pc from Earth.

1. **Introduction**

The search for habitable worlds beyond our Solar System has become one of the most dynamic frontiers in modern astronomy and astrobiology. With the discovery of thousands of exoplanets, it has become possible to assess not only the occurrence of exoplanets but also the potential for life-sustaining conditions. Central to this inquiry is the concept of the habitable zone (HZ)—the region around a star where conditions might allow the presence of liquid water on a planet's surface (e.g. Kasting et al. 1993), considered a fundamental prerequisite for life as we understand it on Earth.

The definition and boundaries of the HZ depend on multiple factors, including stellar spectral type and luminosity, and planetary characteristics such as atmospheric composition and density, and Bond albedo (albedo for short). Recent studies of the Kepler and K2 exoplanet database have enabled the identification and statistical study of the occurrence of Earth- and Super Earth-sized planets within the HZ of various star types, sun like ones (G type) as well as K and F type (Dressing et al. 2013, 2015; Hsu et al. 2019; Kunimoto et al. 2020). Of particular interest to our work are the occurrence studies of cool, M-type dwarfs, which have been performed for Kepler data by Morton and Swift (2014), Dressing & Charbonneau (2015),

Mulders et al. (2015), Gaidos et al. (2016), Yang et al. (2020) and Hsu et al. (2020), and for radial velocity surveys by Bonfils et al. (2013), Sabotta et al. (2021), Ribas et al. (2023) and Kaminski et al. (2025).

This work suggests a new approach to estimate the number of potentially habitable planets in our galactic neighborhood, based on re-analysis of data from the *Kepler* mission (Bryson et al. 2021), and on a recent extended model for the HZ (Wandel 2023; 2025).

The classical delineation of the HZ is based on stellar instellation (irradiated energy per unit area at the planet surface), the total amount of radiation received by a planet from its host star. For G-type stars (similar to the Sun), the HZ is similar to that of the Solar System, which spans approximately from the orbit of Venus to that of Mars. The inner edge of the HZ is set by the Moist Greenhouse limit, where increasing water vapor in the stratosphere leads to enhanced escape of hydrogen and ultimately the loss of surface water before the boiling point is reached. The outer edge is defined by the maximum greenhouse limit, where increasing atmospheric $CO_2$ becomes ineffective at preventing surface water from freezing. These thresholds were originally derived from 1D radiative-convective climate models (Kasting et al., 1993), and have since been refined using more sophisticated 3D General Circulation Models (GCMs; e.g., Yang et al., 2014; Shields et al., 2016; xxx).

To provide a broader empirical perspective, Kopparapu et al. (2013) proposed adjusted HZ boundaries based on geological evidence that early Mars and recent Venus may have once sustained liquid water. These empirical limits suggest a wider zone of potential habitability, highlighting the importance of atmospheric composition and planetary feedback mechanisms. A review by Shields (2019) underscored the model dependence of the HZ concept, e.g. accounting for cloud feedbacks, planetary rotation, obliquity, and other 3D climate effects.

Although GCMs provide high-resolution insights, they are computationally intensive, especially when applied across diverse planetary scenarios. Therefore, faster 1D models continue to play a key role in large-scale exoplanetary surveys and parametric studies (e.g., Kopparapu et al., 2013; Wandel, 2018; Koll et al., 2019).

Special attention has been paid to M-type (red dwarf) stars, which are the most frequent stellar type in the galaxy (e.g. Henry et al. 2018, Golovin et al. 2023) and prime targets for exoplanet habitability surveys (Bonfils et al. 2013; Reiners et al. 2018; Hsu et al. 2019, Ribas et al. 2023). Due to their low luminosities, the HZs around M-dwarfs are much closer to the host star, resulting in shorter orbital periods and deeper transit signals, thus improving the detectability of small planets and their potential biosignature gases via transmission spectroscopy (e.g. Tarter et al. 2007; Rodler and López-Morales, 2014).

However, planets in close-in HZs around M-dwarfs face significant habitability challenges. In their early stages, M dwarfs exhibit high levels of magnetic activity, characterized by extreme ultraviolet radiation and coronal mass ejections, which may drive photo dissociation of water and atmospheric erosion (Sanz-Forcada et al. 2011; Luger & Barnes 2015; Tilley et al., 2019). These effects raise concerns about the long-term retention of atmospheres and volatiles on Earth-sized planets. Despite these risks, recent modeling suggests that limited atmospheric circulation, in combination with initial water inventories, could enable the survival of habitable conditions around M dwarfs (Joshi (2003, AsBio; Yang et al., 2014; Shields, 2016; Wandel & Gale, 2020).

In earlier work (Wandel 2023a; 2025) is has been demonstrated that rocky tidally locked planets can maintain surface and subglacial liquid water in an inverse eyeball configuration (where liquid water or ice are present on the surface only at the night side, off star side of the planet) at significantly larger instellation values than rapidly rotating planets. In other words, this means that such planets can sustain liquid water at smaller distances from the host star, i.e. the effective HZ is significantly extended beyond the inner edge of the conservative HZ. This effect is enhanced for reduced atmospheric and/or oceanic heat transport mechanisms that redistribute stellar energy from the irradiated day side to the night side. Likewise, planets that harbor internal heat sources, such as geothermal or tidal heating, may sustain subsurface oceans beneath thick ice layers even when located beyond the traditional outer boundary of the HZ. These considerations shift the effective inner and outer boundaries of the HZ both physically and in terms of stellar flux (instellation).

## 2. Methods

Like the HZ of the Sun, which extends roughly between the orbits of Venus and Mars, the HZ of other stars depends on the star's luminosity, as the planet's temperature depends on the stellar irradiation, or instellation. For Main Sequence stars the luminosity is related to the stellar type or surface effective temperature, so the location and width of the HZ depend the stellar type of the host (Fig. 1).

The temperature distribution on the surface of locked planets can be described primarily by two parameters: the heating factor *H*, which combines the stellar irradiation with the albedo and the atmospheric impact (e.g. greenhouse effect), and the heat transport parameter (Wandel 2018) *f, which* is the heat fraction at any point on the surface transported by advection and distributed over the entire planet. For example, *f=0.1* means that at any point on the planet surface, 10% of the stellar energy radiated onto the planet is advected and finally distributed equally all over the planet. The thinner the atmosphere and the weaker the global heat transport (f<<1) of a tidally-locked planet, the lower is its night side temperature and the less it is influenced by stellar and atmospheric heating. This shielding may enable the existence of subsurface liquid water on the night side of locked planets much closer to the host star than the inner boundary of the HZ (Wandel 2023a; 2025).

Fig. 1 describes the boundaries in the effective temperature vs. instellation diagram for various HZ-models. The CHZ boundaries (Kopparapu et al. 2013) are denoted by solid green lines, while the extended inner HZ boundary considering liquid water or subglacial water on the night side of locked planets is shown by dashed blue lines.

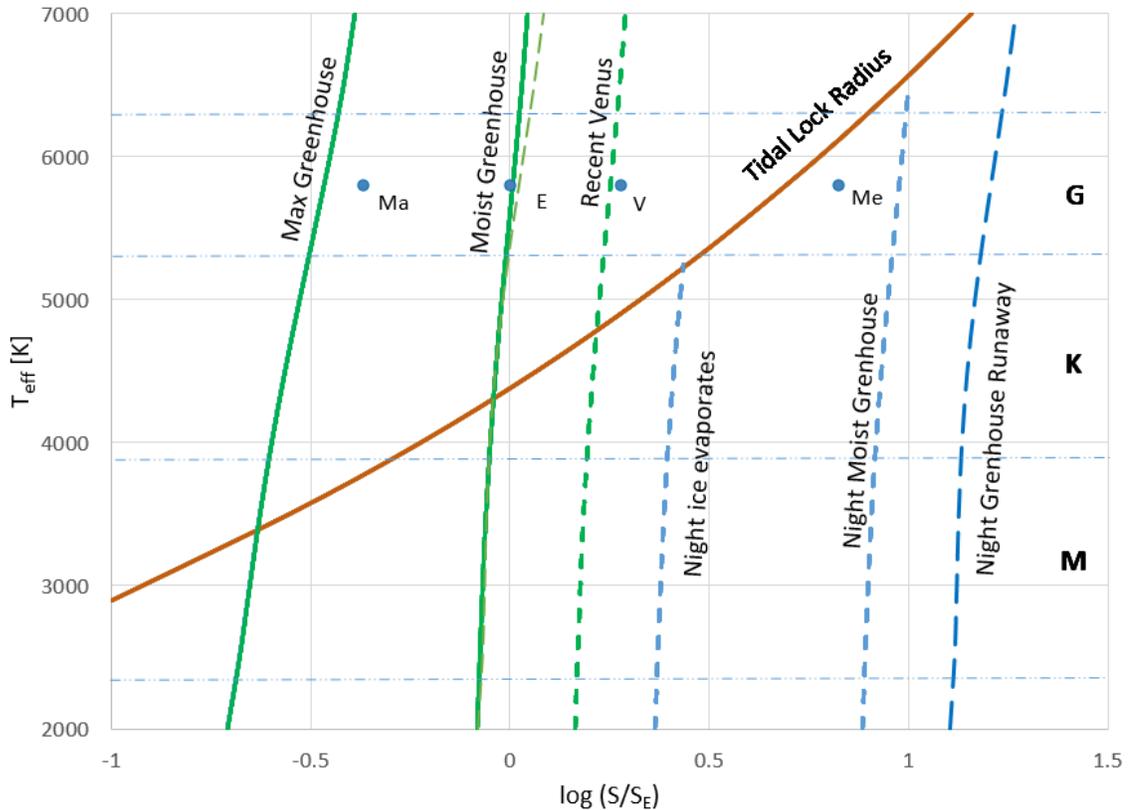

Fig. 1. Boundaries of the Habitable Zone. The horizontal axis marks the radiative flux relative to Earth and the vertical ones, the effective temperature and approximate spectral type of the host star (spectral type boundaries are marked by horizontal thin dashed lines). Solid green curves indicate the conservative HZ boundaries for rapidly rotating planets. Also shown is the Recent Venus inner HZ boundary (green dashed) and the Greenhouse Runaway one (light green dashed curve). The blue dashed curves represent the inner extended HZ boundary for locked planets (Wandel 2025). A global heat transport of 30% ($f$=0.3) and albedo of 0.3 are assumed. The curve marked "Night ice evaporation" is the moist greenhouse curve shifted to the right by a factor of 1/0.3=3.3. The curves marked "Night Moist Greenhouse" and "Night Greenhouse runaway" are the moist and runaway greenhouse curves for $f=0.3$ (Wandel 2025). The red curve marks the tidal locking radius (Wandel 2023b) and blue dots denote Solar System planets.

3. Analysis and results

A broader HZ means more potentially habitable planets per star. Including M dwarfs drastically increases the space density of potentially habitable exoplanets (cf. Kasting et al. 2013). This also enhances the number of planets with a potential for biosignatures in their transmission spectra. In particular, the extension of the HZ for tidally-locked planets of M dwarfs, which are much more abundant than Sun-like stars, leads to a large increase in the number of targets suitable for biosignature detection.

Wandel (2023a) gave an estimate of the number of habitable planets as a function of the distance from Earth (space volume) for G, K and M stars for two values of planet occurrence per star, 0.2 for the Conservative HZ (Kopparapu et al. 2013) and 1.0 for the Extended HZ (Wandel 2023a). Here we elaborate on this calculation by using updated *Kepler* statistics and improved estimates of the HZ-extension for locked planets of K and M stars.

The occurrence of habitable planets depends on the assumed HZ model. We estimate the planet occurrence by integrating over the "phase space", namely the area in the HZ diagram of host star temperature vs. instellation (Fig. 1). In other words, we integrate the planet distribution function over radiative flux received by the planet and the host stellar type (effective temperature), convolving with the space density of the respective stellar types. In this work we limit the analyses to the inner HZ. The outer HZ extensions due to subglacial liquid water and geothermal heat (Wandel 2023a), which is only weakly related to the star's irradiation (Ojha 2022) and the empirical boundary of the Mars polar lakes will be treated in a separate work. Our results give therefore a lower limit on the occurrence of habitable planets. Note that the inner boundaries (dashed blue curves in Fig. 1) terminate at the tidal lock radius, as night ice and night subglacial water require a tidally locked planet; a fast rotating planet at this distance would be too hot to maintain surface ice or liquid water.

Following Bryson et al. (2021), who calculated the occurrence of small planets of main-sequence dwarf stars based on the Kepler DR25 planet candidate catalog and Gaia-based stellar properties, we used these data to calculate the occurrence (per star and spectral type) of planets.

We integrate over instellation onto the planet and host star effective temperature for various HZ models: the conservative HZ (CHZ), between the moist greenhouse and maximum greenhouse boundaries in Fig. 1, the optimistic HZ (OHZ), between the recent Venus and the early Mars (which is close to the maximum greenhouse) boundaries and the extended HZ models.

Bryson et al. (2021) found the occurrence of planets in the radius range 0.5-1.5 Earth radii, orbiting stars with effective temperatures between 4800 K and 6300 K (stellar types from late F to mid K type), at a to be 0.37-0.60 in the CHZ and 0.58-0.88 for the OHZ (roughly flux ranges of *0.2-1.0 and 0.2-2.0 $S_E$*, respectively).

We normalize our integrated HZ-area convolved with the host stellar type to the planet occurrence of Byson et al. (2021). We calculate the HZ area in *S-T*eff plane for their effective temperature range for the CHZ and OHZ models (rows 2- 3 in table 1). We then determine the normalization factor required to match the values of the midpoints of their above occurrence ranges, 0.50 and 0.73 planets per star in the CHZ and OHZ, respectively (column 5 of Table1). Both give the same normalization factor, NF. For example, for the OHZ model:

NF=occurrence/convolved HZ area in *S-Teff* plane x percentage of stars in this effective temperature range=0.73/0.019x13%=5.0. The stellar type occurrence is estimated as follows: 7% (G)+ 5% (early-mid K) + 1% (late F)=13%.

We extrapolate the results to M dwarfs, in the effective temperature range of 2300K-3900K (e.g. Cifuentes et al. 2020). We check the validity of the extrapolation by comparing to Hsu et al (2020).

The HZ phase space area is calculated by integration over the plane of host star $T_{eff}$ and instellation S (Fig. 1):

$$A_{HZ}(\Delta T_{eff}, \Delta S) = \int_{T1}^{T2}[S_{in}(T) - S_{out}(T)]dT, \qquad \text{eq. 1}$$

where the boundaries in $T_{eff}$ (horizontal light dashed lines in Fig. 1) are taken as 2300-3900K for M dwarfs, 3900-5300K for K dwarfs and 5300-6300K for G and late F dwarfs. The functions describing the inner and outer HZ boundaries in the $T_{eff}$-S plane, $S_{in}(T)$ and $S_{out}(T)$, respectively, are taken from Kopparapu et al. (2013, eq. 2), described by the solid green curves in Fig. 1. The extended HZ boundaries (dashed blue curves in Fig. 1) are derived in Wandel (2025).

In order to avoid overweighing early stellar types (high $T_{eff}$) and high $S_E$ we use a logarithmic integration scheme, $dT/T$ and $\log(S)$ rather than $dT$ and $S$, respectively,

$$A_{HZ}(\Delta T_{eff}, \Delta \log S) = \int_{T1}^{T2} \log[S_{in}(T)/S_{out}(T)]dT/T. \qquad \text{eq. 2}$$

To calculate the actual planet occurrence rates we convolve the HZ area in the $S$-$T_{eff}$ diagram with the relative stellar occurrence (fraction of all stars) of G, K and M dwarfs, taken as 7% (Boettner et al. 2024), 13% (Bochanski et al. 2010; Zhu et al. 2025) and 72%, (Golovin et al. 2023) respectively. The remaining 8% are mainly white dwarfs and F stars. Reylé et al. (2021, table 3) gives slightly lower values for the fractions of G and K dwarfs.

The convolved HZ-area is calculated by the scheme

$$\sum_i AHZi(\Delta T_{eff}, \Delta \log S) \; \Phi i(T_{eff}), \qquad \text{eq. 3}$$

where $i$ runs over the relevant star types (e.g. G, K and M) and $\Phi i(T_{eff})$ is the corresponding stellar occurrence (e.g. 7% for G dwarfs).

We normalized our calculations for the occurrence of planets per star for G and K dwarfs to that derived by Bryson et al. (2021) for the conservative HZ (~0.5±0.1 habitable planets per star). Compensating for their narrower effective temperature interval 4800-6300K, our larger interval 3900<T<6300K that includes all K type stars, would give a higher occurrence. we derive the abundances of rocky planets (*0.5-1.5 $R_E$*) for various combinations of host star types and HZ-models; the CHZ and the extended moist greenhouse and runaway greenhouse night side water. For the extended HZ models, we consider only K and M dwarfs, as these models require tidal locking, and as can be seen in Fig. 1, planets of G type stars are not tidally locked in most of this region in the diagram. We calculate the number of potential habitable small planets within a given distance from the Sun (Table 1 and Fig. 2).

| (1) HZ model star-type | (2) HZ area log(T-S) | (3) conv.HZ areax10³ | (4) Fraction of stars | (5) HZ planets /star/type | (6) HZ planets per star | (7) Hab.planets within 10 pc |
|---|---|---|---|---|---|---|
| G type CHZ | 0.102 | 7 | 7% | 0.50 | 0.04 | 14 |
| *G-mid K CHZ (Bryson) | 0.146 | 13 | 13% | 0.49 | 0.07 | 23 |
| *G-mid K OHZ (Bryson) | 0.220 | 19 | 13% | 0.73 | 0.09 | 33 |
| G & K type CHZ | 0.216 | 23 | 20% | 0.57 | 0.12 | 42 |
| *Late K- early M CHZ (Hsu) | 0.119 | 29 | 42% | 0.35 | 0.15 | 51 |
| K type ExtHZ Moist GH | 0.322 | 42 | 13% | 1.62 | 0.21 | 74 |
| *K ExtHZ Runaway GH | 0.370 | 48 | 13% | 1.85 | 0.24 | 84 |
| GKM type CHZ | 0.364 | 129 | 92% | 0.70 | 0.65 | 226 |
| *M Mars/Venus Op.HZ | 0.213 | 149 | 72% | 1.03 | 0.74 | 260 |
| KM Mars/Venus Op.HZ | 0.378 | 175 | 85% | 1.03 | 0.86 | 306 |
| M ExtHZ Moist GH | 0.401 | 289 | 72% | 2.01 | 1.45 | 505 |
| *KM ExtHZ Moist GH | 0.722 | 330 | 85% | 1.94 | 1.75 | 612 |
| KM ExtHZ Runaway GH | 0.828 | 378 | 85% | 2.22 | 1.89 | 662 |

Table 1. Abundance of habitable planets for various combinations of host star type and HZ model.

The HZ-model and star type combination in Column 1 of table 1 refer to the curves labelled by the acronyms in the legend of Fig. 2 (excluding the ones marked by an asterisk). GH is short for Green House. Column (2): area in the $S$-$T_{eff}$ HZ diagram (eq. 2) using log ($T_{eff}$/ 5780K) and log ($S/S_E$), respectively). Column (4): the area in the S-$T_{eff}$ diagram convolved with the respective fractions (eq. 3) of G, K and M-dwarfs, or their combinations. It is calculated by dividing column 2 by column 3 and normalizing to 0.5 for the first row ( G CHZ). Column (5): average number of habitable planets per star considering only the type(s) indicated in the first column. Column (6): average number of habitable planets per star considering all stars. Column (7): number of habitable planets within 10 pc from the Sun. For the extended HZ models, the global heat transport has been assumed to be $f$=0.3. The inaccuracy of our integration procedure is roughly 30%, which is much less than the intrinsic uncertainties, further discussed in the next section.

Column 7 in Table 1 uses the work of Reylé et al. (2021), who find 350 stars within 10 pc, excluding sub stellar objects (later than mid-L type), but including 41 stars with no identified type (assumed by the authors to be mostly M dwarfs) and 20 white dwarfs, which gives a local star number density of 0.09±0.01. Including all objects except exoplanets their data give 0.11 stars pc$^{-3}$. Golovin et al. (2023) gives from the CNS5 catalogue 0.08±0.01 stars pc$^{-3}$, 5% of which are white dwarfs.

We compare our occurrence rate of planets in the CHZ of M-dwarfs to the results of other works which use transit (Kepler) and radial velocity surveys.

For the Kepler data Hsu et al. (2020) find 0.33±0.11 habitable planets per star for planets with 0.75-1.5 $R_E$ (or $R_⊕$ ) size in the range early-type M dwarfs and late K dwarfs. In order to compare our scheme to their work, we calculated the occurrence of planets within the CHZ for the same effective temperature range (Late K-early M). We find 0.35±0.10 planets per star (Table 1, the row marked "Late K-early M CHZ (Hsu)"), a good agreement which supports the extrapolation of the normalization from Byson's FGK sample to M dwarfs.

For radial motion we compare with Bonfils et al. (2013). Analyzing a 6-yr radial velocity study of 102 nearby southern M dwarfs, they found 0.41−0.13+0.54 rocky Superearth planets ($M \sin i$=1-10 $M_E$) per star in the optimistic HZ (recent Venus/ early Mars HZ boundaries). In order to compare our scheme with the result of Bonfils et al. (2013) we calculated the occurrence per star of planets of M dwarfs within the optimistic HZ boundaries, yielding 0.74±0.22 (Table 1, "M Mars/Venus Op.HZ"). While the two values agree within the uncertainties, the lower occurrence found by Bonfils et al. (2013) makes sense, as they only include planets in the mass range of $M \sin i$=1-10 $M_E$, (see their table 11 for extension to other mass ranges) while our occurrence calculations (normalized to the Kepler survey) include all planet masses.

More recent RV-analyses are given by Sabatto et al. (2021) and Ribas et al. (2023), who considered all planets, not only habitable. The later analyzed the CARMENES sample of 238 M-dwarf targets with planet masses in the range 1 - 1000 $M_E$, finding an occurrence rate of 1.44 ± 0.20 planets per star. As expected this is higher than our result, since it includes all planets, not only the habitable ones (cf. their Fig. 9). Limiting the planet mass to the Superearth range ($M \sin i$=1-10 $M_E$) Ribas et al. (2023) find a lower occurrence, 1.05±0.15. Comparison with other works, RV as well as transit surveys (their Fig. 12) shows a scatter of a factor of ~2, which gives an idea of the overall error range of the occurrence results.

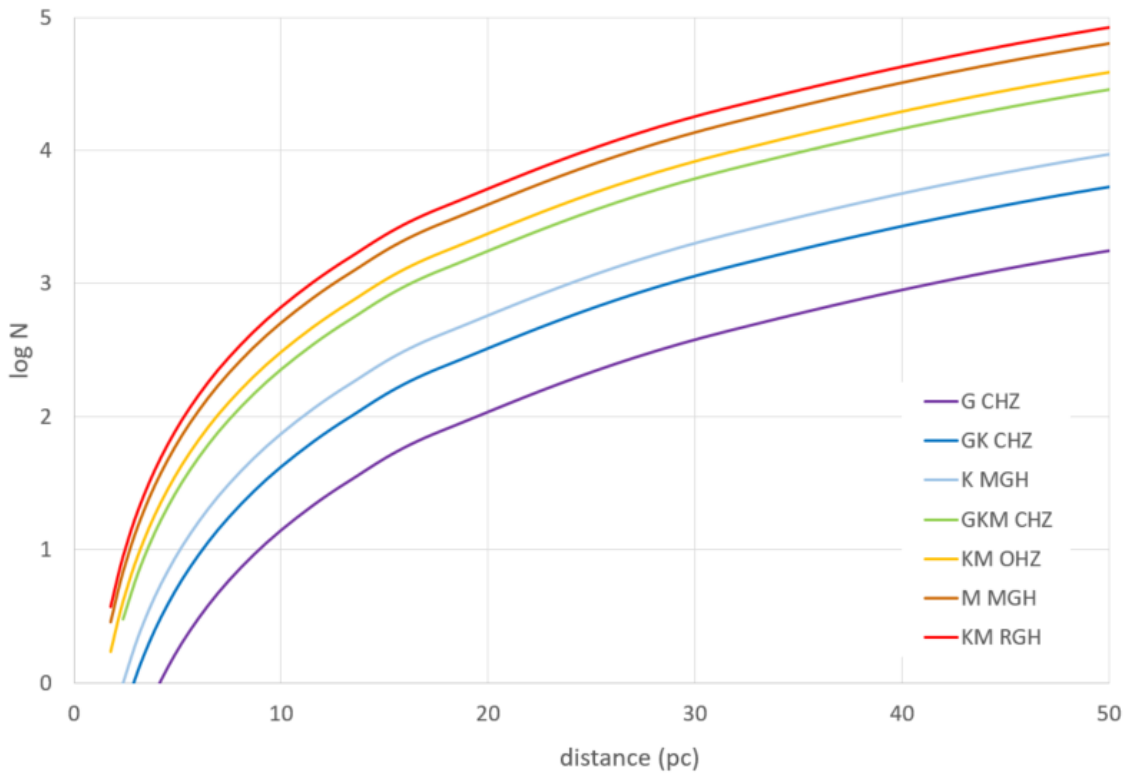

Fig. 2. The number of expected habitable planets as a function of the distance from the Sun for various combinations of host star type and HZ-models, corresponding to Table 1. The lowest curve (top one in the legend) is for the conservative HZ (CHZ), with G dwarfs only. The next one includes also K dwarfs and the forth one – G, K and M dwarfs. The curves marked by MGH in the legend are for the Moist Green House limit on the night side of locked planes. RGH stands for the Runaway Green House limit on the night side, while OHZ is for the optimistic HZ (Early Mars-Recent Venus). As in table 1, $f$=0.3 is assumed for all extended HZ models (MGH and RGH).

## 4. Conclusions and final remarks

Extending the classical boundaries of the conservative HZ - traditionally defined based on the ability of a planet to sustain surface liquid water under Earth-like atmospheric conditions - can significantly broaden the scope of potentially habitable exoplanets. Incorporating additional planetary scenarios such as tidally locked rotation and subsurface liquid water on the night side of locked planets can improve our understanding of planetary habitability and its distribution in our galactic neighborhood.

Using earlier results (Wandel 2023a, 2025), that tidally locked planets can maintain stable regions of surface liquid water on their night sides, even when located much closer to their host stars than previously considered habitable.

A lower heat transport would increase these numbers. For example, if $f$ were equal to 0.1 rather than 0.3, the curves for the extended HZ in Fig. 1 would move by log 3~0.5 units to the right, with a corresponding increase in the HZ area in the diagram and number of planets.

Consequently, the numbers derived in this work may be a conservative estimate. If global heat transport mechanisms are less efficient, or if additional factors - such as thin atmospheres, or specialized planetary climates (e.g., an eyeball climate) - are taken into account, the potential occurrence or abundance of habitable planets could increase substantially.

Taking into account the low-instellation HZ-boundary (beyond the outer geometrical HZ-boundary) would further enhance the number of potentially habitable planets. These findings underscore the importance of broadening the criteria used in future exoplanet surveys and theoretical models of habitability, allowing for a more inclusive and realistic assessment of the prevalence of life-supporting environments.

By translating these physical HZ extensions into flux-based boundaries and integrating across a multidimensional phase space - specifically the host star's spectral type and instellation - it becomes possible to estimate the statistical occurrence of habitable planets for various stellar populations. Our modeling suggests that, assuming a moderate (e.g., ~30%) efficiency of global heat redistribution, the expanded HZ could support the existence of approximately one to two small habitable planets per star, which gives an estimated population of up to 700 potentially habitable exoplanets within 10 parsecs of the Sun.

There are several specific uncertainties in the occurrences calculated, such as the relation between the HZ "phase space" and the occurrence of habitable planets e.g. by semi major axis or instellation, or the appropriate integration procedure, e.g. whether simple (eq. 1) or logarithmic (eq. 2).

More general, deeper questions are the validity of the normalization to the occurrence of CHZ planets of G and K dwarfs (Bryson at a. 2021) and of the extrapolation from G and K host types to M dwarfs. to the results of may be tested by comparing to other results on exoplanet occurrence rate for M-dwarfs with the conservative HZ. We may gain more confidence by comparing our results for planets in the CHZ of M dwarfs to earlier works.

The logarithmic scale for the integration of the extended HZ in the habitability space may underestimate the number of planets compared with integrations over a linear scale (either of instellation or physical distance). Bryson et al. used a linear scale, integrating over *dS* rather than over *d(log S)*, as done in this work. We consider the logarithmic scale more appropriate for our case, as it gives more weight to the inwards extended HZ, i.e., higher instellations closer to the host star, which are those preferentially sampled by the currently available transit and transmission spectroscopic data.

Comparing column 6 in Table 1 with the *total* number of stars within *10* pc from the Solar System, we conclude considering planets in the extended HZ of M-dwarfs there is more than one habitable planet per star.

**Acknowledgements:** this research has been supported by the Minerva Foundation, at the Center for Studying the Planetary Emergence of Life. Important suggestions of the reviewer helped to significantly improve this work.